\documentclass[a4paper,11pt]{article}
\pdfoutput=1 
\usepackage[utf8]{inputenc}
\usepackage{jheppub}
\usepackage{graphicx}
\usepackage{caption}
\usepackage{subcaption}
\usepackage{hyperref}
\usepackage[T1]{fontenc}

\usepackage{breqn}

\newcommand{\be}{\begin{equation}}
\newcommand{\ee}{\end{equation}}
\newcommand{\bea}{\begin{eqnarray}}
\newcommand{\eea}{\end{eqnarray}}

\newcommand{\nn}{\nonumber\\}
\def\CO{\mathcal{O}}

\def\qfr{\mathfrak{q}}
\def\wfr{\mathfrak{w}}

\title{Adding new branches to the "Christmas tree" of the quasinormal spectrum of black branes}

\author[a]{Sa\v{s}o Grozdanov}
\author[b]{and Andrei O. Starinets}
\affiliation[a]{Center for Theoretical Physics, Massachusetts Institute of Technology, \\ Cambridge, MA 02139, USA}
\affiliation[b]{Rudolf Peierls Centre for Theoretical Physics, University of Oxford, \\ Parks Road,  Oxford OX1 3PU, 
UK}
\emailAdd{saso@mit.edu}
\emailAdd{andrei.starinets@physics.ox.ac.uk}

\abstract{In holography, quasinormal spectra of black branes coincide with the poles of retarded finite-temperature correlation functions of a dual quantum field theory in the limit of infinite number of relevant degrees of freedom such as colours. For asymptotically anti-de Sitter backgrounds, the spectra form a characteristic pattern in the complex frequency plane, colloquially known as the ``Christmas tree''. At infinite coupling, the tree has only one pair of branches. At large but finite coupling, the branches become more dense and lift up towards the real axis, consistent with the expectation of forming a branch cut in the limit of zero coupling. However, it is known that at zero coupling, the corresponding correlators generically have not one but multiple branch cuts separated by intervals proportional to the Matsubara frequency. This suggests the existence of multiple branches of the ``Christmas tree'' spectrum in dual gravity. In this note, we show numerically how these additional branches of the spectrum can emerge from the dual gravitational action  with higher-derivative terms. This phenomenon appears to be robust, yet, reproducing the expected weak coupling behaviour of the correlators quantitatively implies the existence of  certain constraints on the coefficients of the higher-derivative terms of the dual gravity theory.}

\preprint{MIT-CTP/5092, OUTP-18-10P}

\begin{document} 
\maketitle
\flushbottom

\section{Introduction}
\label{sec:intro}
Quasinormal spectra of black branes play an important role in studies of strongly interacting 
 thermal quantum field theories (QFTs) via gauge-gravity duality methods. In the early days of the duality exploration, it was pointed out that  small fluctuations of  Einstein-Hilbert gravity backgrounds with event horizons should correspond to thermalisation processes in the dual QFTs at infinitely strong coupling  \cite{KalyanaRama:1999zj},  \cite{Horowitz:1999jd}, and in fact it was noticed that the spectrum of the BTZ black brane coincides with the location of the poles of the retarded correlators in a dual (infinitely strongly coupled) two-dimensional CFT \cite{Birmingham:2001pj}. With the recipe for computing the Minkowski space correlators from dual gravity established, this connection was shown to be a generic consequence of the gauge-string duality conjecture \cite{Son:2002sd}. This was further explored in refs.~\cite{Starinets:2002br,Nunez:2003eq,Kovtun:2005ev} and subsequent publications \cite{Berti:2009kk}. 
 
\begin{figure}[htbp]
\centering
\includegraphics[width=0.45\textwidth]{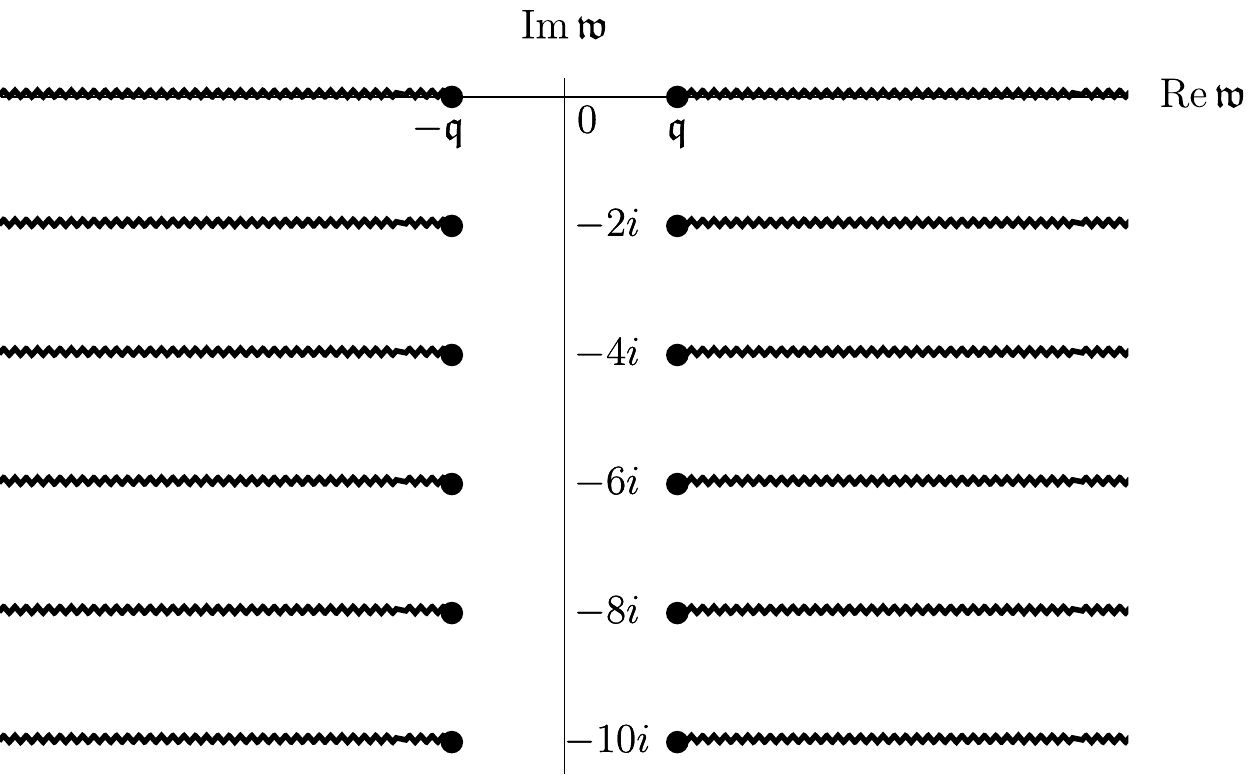}
\includegraphics[width=0.45\textwidth]{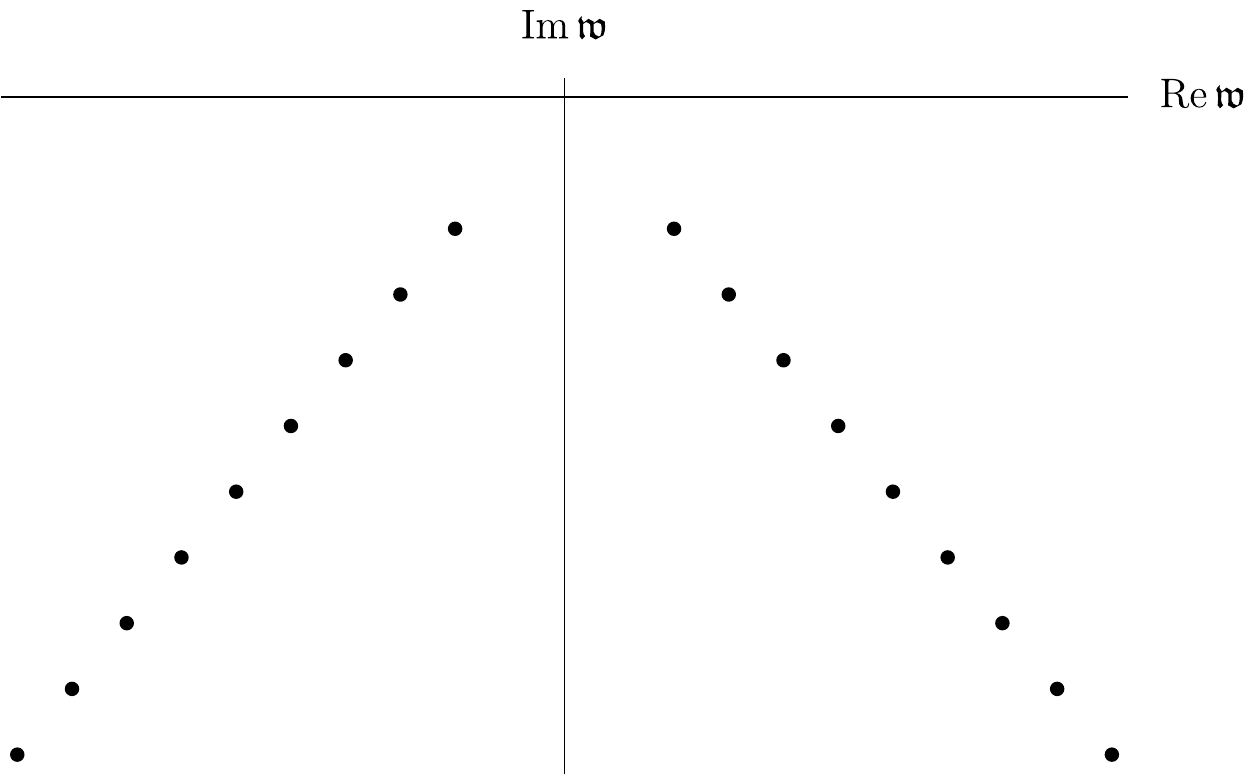}
\caption{Singularities of a thermal retarded two-point  function $G^R(\wfr,\qfr)$ at $\qfr\neq 0$ in the complex frequency plane at zero coupling \cite{Hartnoll:2005ju} (left panel) and infinitely large coupling \cite{Starinets:2002br} (right panel).}
\label{fig:cuts_poles}
\end{figure}

 Quasinormal modes of black branes in five dimensions, dual to four-dimensional thermal QFTs, typically obey differential equations of the Heun type \cite{Starinets:2002br}, 
 whose generic analytic solution is unknown.  A rare analytic example in $4d$ is provided by 
 the  retarded two-point correlator function $G^R_{JJ}(\omega,q)$ of the R-currents in ${\cal N}=4$ supersymmetric $SU(N_c)$ Yang-Mills (SYM) theory in the limit of  infinite $N_c$ and infinite `t Hooft 
 coupling $\lambda=g^2_{YM}N_c$. At zero spatial momentum, all components of the correlator are proportional to \cite{Myers:2007we}
\begin{align}
G^R_{JJ}(\wfr,\qfr=0) \sim \Pi (\wfr) = \frac{N_c^2 T^2}{8} \left\{ i \wfr +
 \wfr^2 \left[ \psi \left( \frac{(1-i)\wfr}{2}\right) + \psi \left( - \frac{(1+i)\wfr}{2}\right)\right] \right\}\,,
 \label{r-current-correlator}
\end{align}
 where $\wfr = \omega/2\pi T$, $\qfr =q/2\pi T$, and $\psi (z)$ is the logarithmic derivative of the Gamma function. The retarded correlator \eqref{r-current-correlator} is a meromorphic function in the complex $\wfr-$plane, with the poles located at
\begin{align}
\wfr_n =  \left(\pm 1-i \right) n\,, \qquad n=1,2,3, \ldots \,.
\end{align}
 These are precisely the quasinormal modes of the dual AdS-Schwarzschild black brane background \cite{Nunez:2003eq}. At $\qfr\neq 0$, the quasinormal spectra with $\wfr_n=\wfr_n(\qfr)$ of this and other correlators, now found numerically, have
  a similar pattern in the complex $\wfr-$plane, colloquially known as the ``Christmas tree'' (see Fig.~\ref{fig:cuts_poles}). 

At zero `t Hooft coupling, the structure of singularities of retarded thermal correlators is very different: at $\qfr\neq 0$, they are branch points\footnote{Correlators at $\qfr=0$ are considered in section \ref{sec:zero-q}.} rather than 
poles\footnote{Analytic structure of thermal correlators at small but finite coupling is poorly understood. For 
recent work in this direction, see refs.~ \cite{Moore:2018mma,Romatschke:2015gic, Kurkela:2017xis,Grozdanov:2018atb}.}
 \cite{LGY}, \cite{Hartnoll:2005ju} (see  Fig.~\ref{fig:cuts_poles}). In position space, this corresponds to the  qualitatively different late time dependence of the correlators at weak coupling in comparison to strong coupling. When the coupling is increased from zero (decreased from infinity), we expect the singularity structures shown in the left and right panels of Fig.~\ref{fig:cuts_poles} to transform into each other, respectively. On the strong coupling side, lowering the coupling from infinity corresponds to taking into account string theory $\alpha'$ corrections to the dual gravity background. However, it was argued in ref.~\cite{Hartnoll:2005ju} that {\it perturbative} $\alpha'$ corrections cannot alter the character of the quasinormal spectrum qualitatively: the singularities remain isolated poles, and the branch points do not appear. Numerical investigations in ref.~\cite{Stricker:2013lma}  of the singularities of  thermal energy-momentum tensor correlator via dual gravity with higher-derivative terms treated perturbatively were consistent with those expectations: the poles in the right panel of Fig.~\ref{fig:cuts_poles} just move up slightly and non-uniformly\footnote{Within the perturbative approach of ref.~\cite{Stricker:2013lma}, the poles lying deeper in the complex plane are affected by the 
perturbation stronger than the poles located closer to the real axis. This picture is altered by the ``resummation'' \cite{Waeber:2015oka}.}
  under the influence of the higher-derivative perturbation.

Perturbative treatment of higher-derivative terms in dual gravity is a necessity for two reasons. First, only a few such structures out of an infinite series of corrections in increasing powers of $\alpha'$ are known explicitly in the low energy string theory actions. Thus, any result beyond an infinitesimally small leading-order correction is potentially affected by the contributions coming from the unknown terms. Second, theories with higher derivatives, if treated non-perturbatively, generically suffer from the Ostrogradsky instability and related pathologies. The second problem can be alleviated, at least at first sight, for Lovelock gravity (in particular, for Gauss-Bonnet gravity), where the equations of motion are of second order.   
\begin{figure}[htbp]
\centering
\includegraphics[width=0.45\textwidth]{cuts-zero-coupling.pdf}
\includegraphics[width=0.45\textwidth]{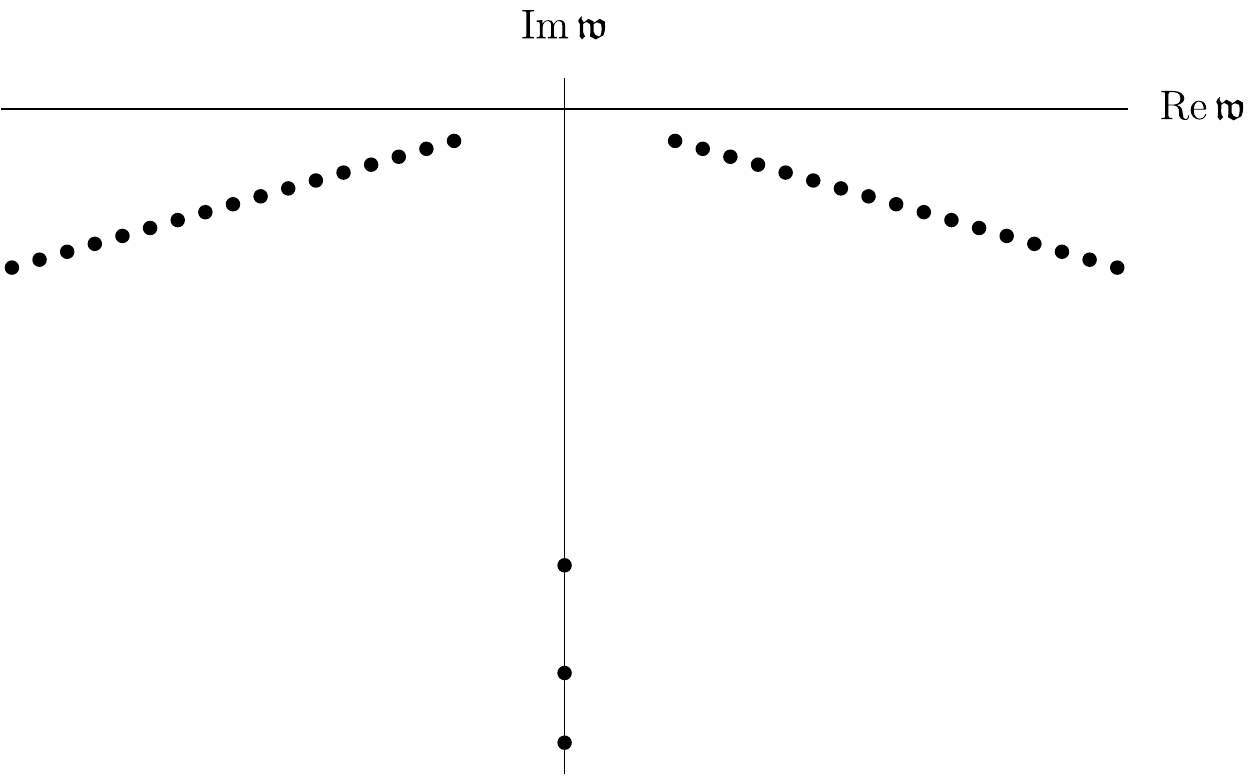}
\caption{Singularities of a thermal retarded two-point  function $G^R(\wfr,\qfr)$ at $\qfr\neq 0$ in the complex frequency plane at zero coupling \cite{Hartnoll:2005ju} (left panel) and  large but finite coupling  \cite{Grozdanov:2016vgg} (right panel).}
\label{fig:cuts_poles_finite}
\end{figure}
The first problem cannot disappear but can be perceived as less acute if one is only interested in a qualitative pattern of behaviour, provided such a pattern is stable. In this case, equations of motion are reduced to second-order ones perturbatively in some relevant small dimensionless parameter $\gamma$ but their solution is found without assuming $\gamma \ll 1$. Such a ``partial resummation'' \cite{Waeber:2015oka} cannot be fully quantitatively correct beyond the limit of small $\gamma$ as it misses potential contributions from terms of order $\gamma^2$ and higher in the action, and those contributions may be significant \cite{Buchel:2018eax}, but 
qualitatively, the emerging pattern of the spectrum appears to be robust. Crucially, this approach allows to uncover parts of the spectrum hidden at complex infinity in the limit $\gamma\rightarrow 0$ \cite{Grozdanov:2016vgg}. This is very well 
illustrated by considering the spectrum of Gauss-Bonnet black branes, where perturbative and non-perturbative regimes can be reliably compared \cite{Grozdanov:2016fkt}. 

The analysis of the quasinormal spectra of black branes in $5d$ Gauss-Bonnet gravity and $\alpha'$-corrected type IIB supergravity \cite{Grozdanov:2016vgg,Grozdanov:2016fkt} revealed a generic qualitative pattern for the behaviour of the thermal spectrum shown in Fig.~\ref{fig:cuts_poles_finite} (right panel). At large but finite coupling (small but finite $\gamma$), the two branches of poles lift up towards the real axis; the real parts of the top-most poles approach the values of $\pm \qfr$; the poles of the correlator also become more dense and interlaced with zeros (this last feature is not shown in Fig.~\ref{fig:cuts_poles_finite}) which appears to be a characteristic feature consistent with the formation of a branch cut \cite{Moore:2018mma}. In addition, there are extra poles on the imaginary axis coming up from below as the coupling is lowered from infinity \cite{Grozdanov:2016vgg}. One may be prepared to believe that with the coupling decreasing all the way to zero,  the dense sequence of poles in the right panel of  Fig.~\ref{fig:cuts_poles_finite} gradually approaches the top branch cut shown in the left panel. But where do the other cuts  in the left panel of  Fig.~\ref{fig:cuts_poles_finite}  come from? And where do the poles on the imaginary axis go then? Similar questions exist for correlators at $\qfr=0$ (see section \ref{sec:zero-q}).
 
 In this note, we show that the missing cuts can come from adding more higher-derivative terms to the gravity action. We find that at large but finite coupling, the standard ``Christmas tree'' of the quasinormal spectrum has more than just one pair of branches (in fact, we expect an infinite number of branches to appear with the full series of higher-derivative terms taken into account but as a proof of principle, we only consider one additional pair of branches in this paper). The new branches of the ``Christmas tree'' emerge as a result of complicated dynamics and collisions of poles on the imaginary axis. The phenomenon of two poles colliding and splitting off the imaginary axis is known in holography, mostly in the context of models with finite chemical potential and finite temperature (see e.g. \cite{Davison:2011ek,Grozdanov:2018ewh,Gushterov:2018spg,Gushterov:2018nht}). The same phenomenon is responsible for the breakdown of the hydrodynamic description at sufficiently large, coupling-dependent spatial momentum \cite{Grozdanov:2016vgg,Grozdanov:2016fkt} (see also \cite{Grozdanov:2016zjj,DiNunno:2017obv}).\footnote{See ref. \cite{Grozdanov:2018fic} for a discussion of the connection between the phenomenon of colliding poles in the quasihydrodynamic regime and the existence of approximately conserved symmetries.} Here, this mechanism leads to the formation of the extra branches of the ``Christmas tree'' spectrum, as we describe below.

\section{New branches of the quasinormal spectrum in higher-derivative gravity}

We consider  the bulk $5d$ gravity action with higher-derivative terms organised as a power series of 
the Riemann tensor squared, truncated at second order,
\begin{align}\label{ActR2R4}
S = \frac{1}{2\kappa_5^2} \int d^5 x \sqrt{-g} \left[ R - 2 \Lambda - \alpha \gamma_1 R_{\mu\nu\rho\sigma} R^{\mu\nu\rho\sigma} -  \alpha^2 \gamma_2 \left(R_{\mu\nu\rho\sigma} R^{\mu\nu\rho\sigma}\right)^2  \right],\,
\end{align}
where $\left|\alpha\right| \ll 1$ is a book-keeping parameter reminiscent of $\alpha'$ in string theory. This choice of the action is partially motivated by simplicity and partially by experience with field redefinitions and their role in computing thermodynamic and transport coefficients in theories with higher-derivative terms \cite{Gubser:1998nz,Brigante:2007nu,Grozdanov:2015asa,Grozdanov:2014kva}. All dimensionful quantities will be measured in units set by the cosmological constant  $\Lambda = - 6 / L^2$. Without loss of generality, we set $\gamma_1 = 1$ and $L=1$ in the following, unless noted otherwise. 

The equations of motion corresponding to the action  \eqref{ActR2R4} possess a black brane background solution with the metric of the form
\begin{align}
ds^2 = - f(r) dt^2 + \frac{dr^2}{h(r)} + r^2 \left(dx^2 + dy^2 + dz^2 \right)\,,
\label{bb-metric}
\end{align}
where $f(r)$ and $h(r)$ can be found perturbatively in the small $\alpha$ expansion:
\begin{align}
f(r) &= f_0 (r)+ \alpha f_1 (r) + \alpha^2 f_2(r) , \label{bg1} \\
h(r) &= f_0 (r)+ \alpha h_1 (r) + \alpha^2 h_2(r) . \label{bg2}
\end{align}
The functions entering eqs.~\eqref{bg1} and \eqref{bg2} are given by
\begin{align}
f_0(r) &= r^2 \left ( 1 - \frac{r^4_0}{r^4} \right) , \\
f_1(r) &= \frac{2 r_0^4}{r^4} f_0(r), \\
h_1(r) &= \left( -\frac{2}{3} + \frac{2 r_0^4}{r^4}\right) f_0 (r),\\
f_2(r) &= \frac{4 r_0^4 }{3 r^{4}} \left[  (1164 \gamma_2+149) +6 (174 \gamma_2 -7)  \frac{ r_0^4}{r^8} -972 \gamma_2 \frac{r_0^8}{r^8} \right]    f_0(r) ,\\
h_2(r) &= \frac{8}{9} \left[  (90 \gamma_2 +1) +6 (291 \gamma_2 +37) \frac{r_0^4}{r^4} +27 (202 \gamma_2 -17)  \frac{r_0^8}{r^8}-9234 \gamma_2 \frac{r_0^{12}}{r^{12}}\right] f_0(r) , 
\end{align}
where $r_0$ denotes the position of the event horizon. The Hawking temperature of the black brane is
\begin{align}
T= \frac{r_0}{\pi }  \left[1 + \frac{5  }{3} \alpha -  \left(40 \gamma_2 +\frac{605}{18}\right)\alpha ^2\right].
\end{align}

We are interested in computing the quasinormal spectrum of gravitational fluctuations of the background \eqref{bb-metric} and studying its dependence on the parameters of the action. We follow the standard procedure of decomposing the metric fluctuations $h_{\mu\nu}$ into symmetry classes \cite{Son:2002sd,Policastro:2002se,
Kovtun:2005ev,Grozdanov:2016vgg} and, for simplicity, focus on the scalar channel fluctuations $h_{xy}$. 
To order $\CO(\alpha^2)$, and with the higher-derivative terms eliminated 
at the expense of generating $\CO(\alpha^3)$ terms as explained in refs.~\cite{Grozdanov:2016vgg,Solana:2018pbk}, the equation of motion for the metric fluctuation component  $h_{xy}(r) = r^2 Z(r)$, represented 
 in terms of Fourier modes $\sim e^{-i\omega t+ i q z}$, can be written as
\begin{align}\label{ScalarEq}
 Z''(u) + A\,  Z'(u) + B\, Z(u) = 0 ,
\end{align}
where $u = r_0^2/r^2$, $ A = A_0 +\alpha A_1 + \alpha^2 A_2$ and $B = B_0 + \alpha B_1 + \alpha^2 B_2$. The functions appearing in the equation \eqref{ScalarEq} are given by
\begin{align}
A_0(u) &= -\frac{u^2+1}{u-u^3} , \\
A_1(u) &= 12 u ,\\
A_2 (u) &= -8 u \left[12 \gamma_2 \left(261 u^4+278 u^2-39\right)+437 u^2-192 \qfr^2 u-179\right],
\end{align}
and
\begin{align}
B_0(u) =& \,\, \frac{\qfr^2 \left(u^2-1\right)+\wfr ^2}{u \left(u^2-1\right)^2} , \\
B_1(u) =&\,\, \frac{\qfr^2 \left(4-18 u^2\right)-4 \wfr ^2}{u \left(u^2-1\right)} ,\\
B_2 (u) =& -\frac{4 \qfr^2 \left[12 \gamma_2 \left(2799 u^6-2289 u^4+177 u^2+10\right)-2829 u^4+1275 u^2+47\right]}{3 u \left(u^2-1\right)} \nn
& -\frac{4 \wfr ^2 \left[24 \gamma_2 \left(999 u^4-102 u^2-5\right)-837 u^2-47\right]}{3 u \left(u^2-1\right)}.
\end{align}

\begin{figure}[htbp]
\centering
\includegraphics[width=0.45\textwidth]{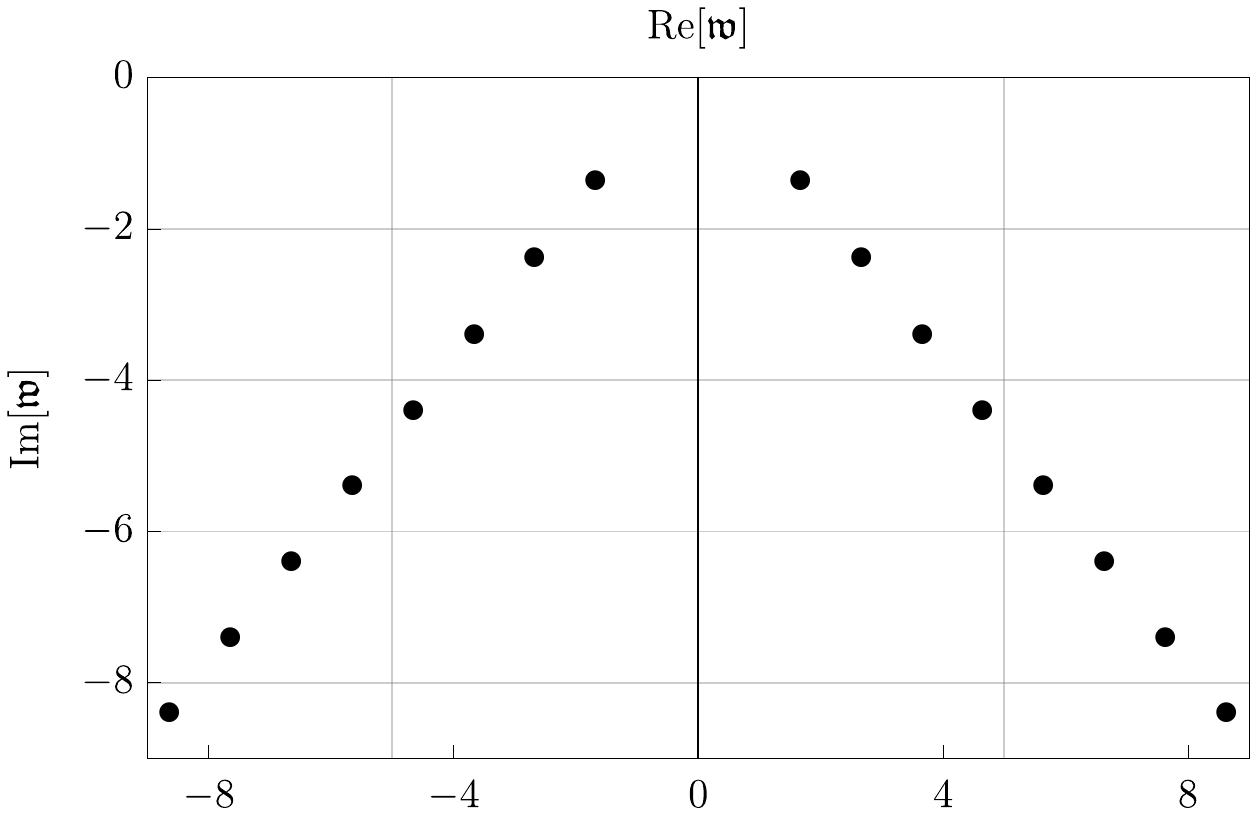}
\hspace{0.05\textwidth}
\includegraphics[width=0.45\textwidth]{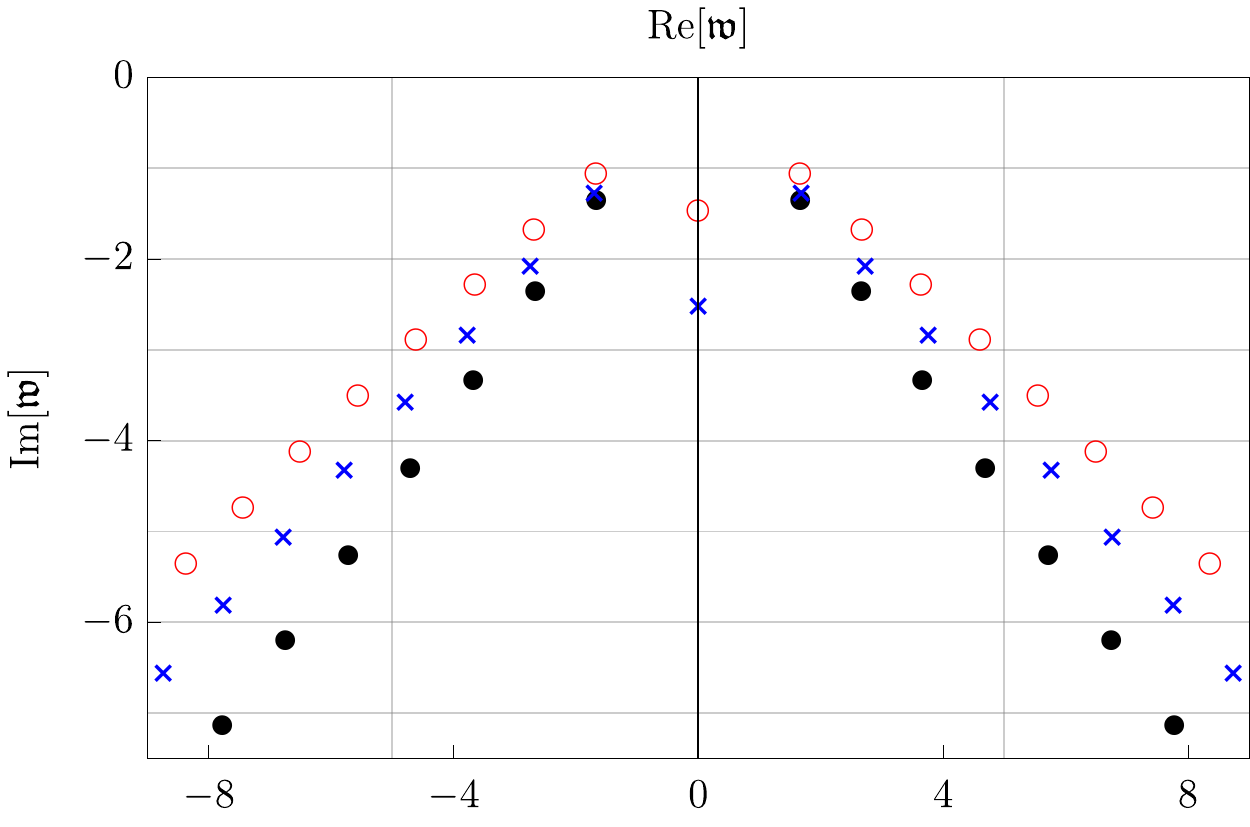}
\caption{The scalar channel quasinormal spectrum at $\alpha=0$, $\qfr=0.5$ (left panel) and in the theory truncated at $O(\alpha)$, i.e. at $\alpha\neq 0$, $\gamma_2=0$, $\qfr=0.5$  (right panel),  plotted for $\alpha = 0.01$ (black dots), $\alpha = 0.05$ (blue crosses) and $\alpha = 0.1$ (red circles).}
\label{fig:orders-0-1}
\end{figure}

At $\alpha=0$, the quasinormal spectrum $\wfr_n$ of eq.~\eqref{ScalarEq} 
(shown in the left panel of Fig.~\ref{fig:orders-0-1}) coincides with the known spectrum 
of the AdS-Schwarzschild black brane \cite{Starinets:2002br}. Analogously, at $\alpha\neq 0$ and $\gamma_2=0$, we recover the spectrum qualitatively similar to the one
 found for black branes in Gauss-Bonnet gravity and AdS-Schwarzschild background corrected by $R^4$ 
 terms \cite{Grozdanov:2016vgg,Grozdanov:2016fkt,Solana:2018pbk}. This spectrum is shown in the right
  panel of Fig.~\ref{fig:orders-0-1} for several values of $\alpha$. Note the presence of the new modes on the imaginary axis, as described in the Introduction. The scalar channel has no hydrodynamic modes on the imaginary axis (or elsewhere), and the new poles can move up unobstructed. The same modes are responsible for the transport peak in the spectral function of a dual thermal QFT at  large but finite coupling \cite{Solana:2018pbk}. 
%

\begin{figure*}[h]
\centering
\includegraphics[width=0.45\textwidth]{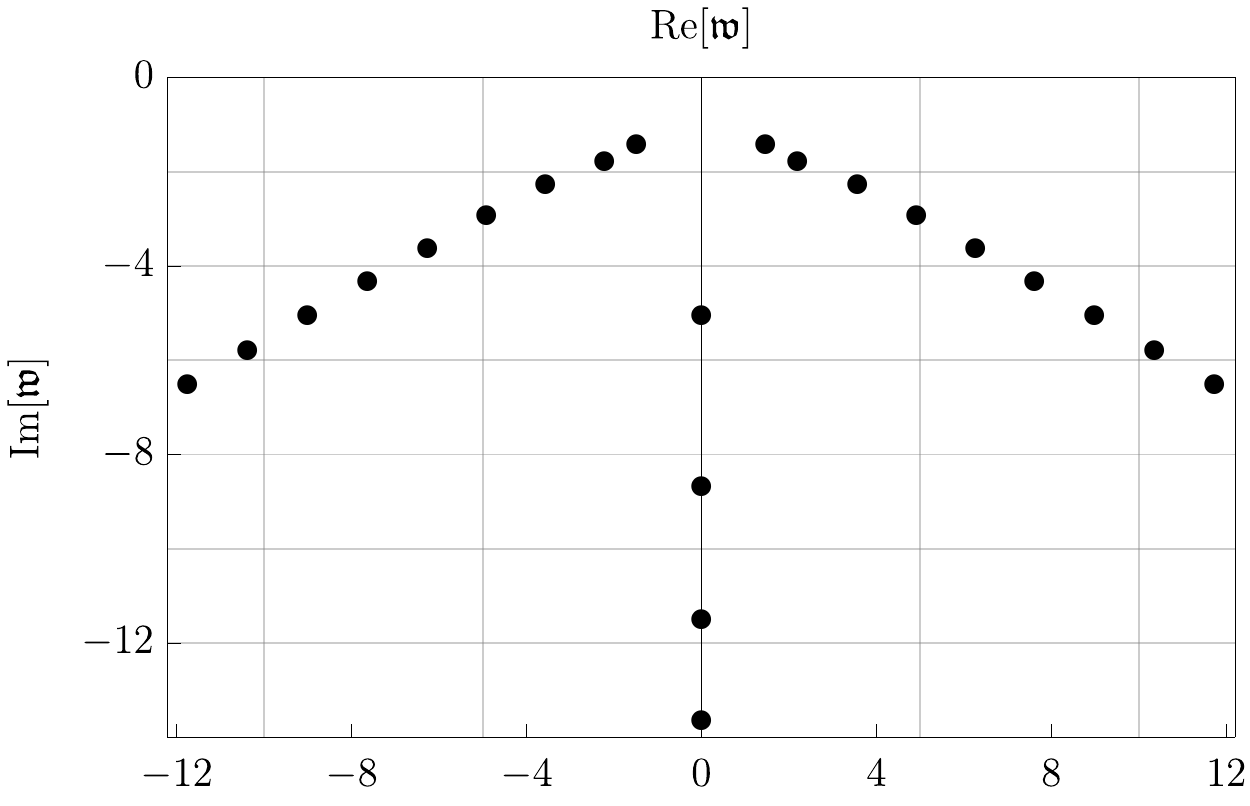}
\hspace{0.05\textwidth}
\includegraphics[width=0.45\textwidth]{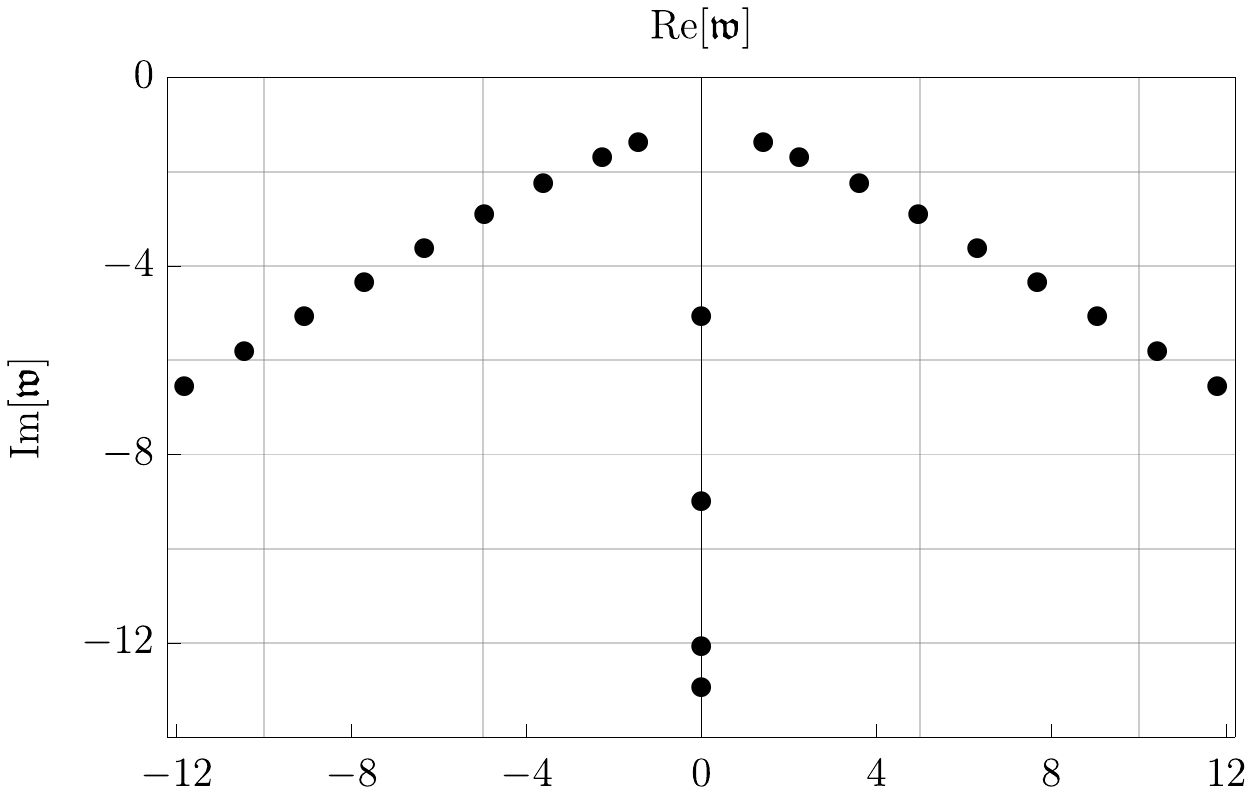}
\\
\includegraphics[width=0.45\textwidth]{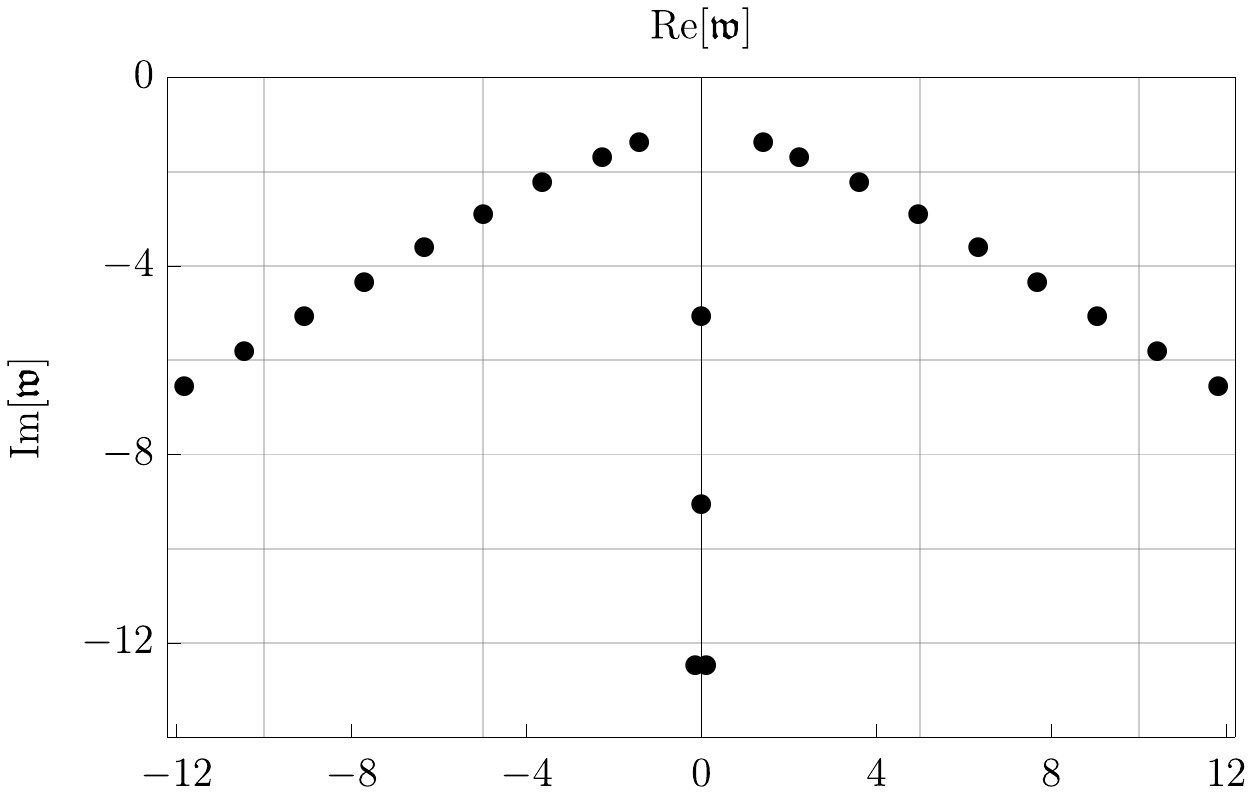}
\hspace{0.05\textwidth}
\includegraphics[width=0.45\textwidth]{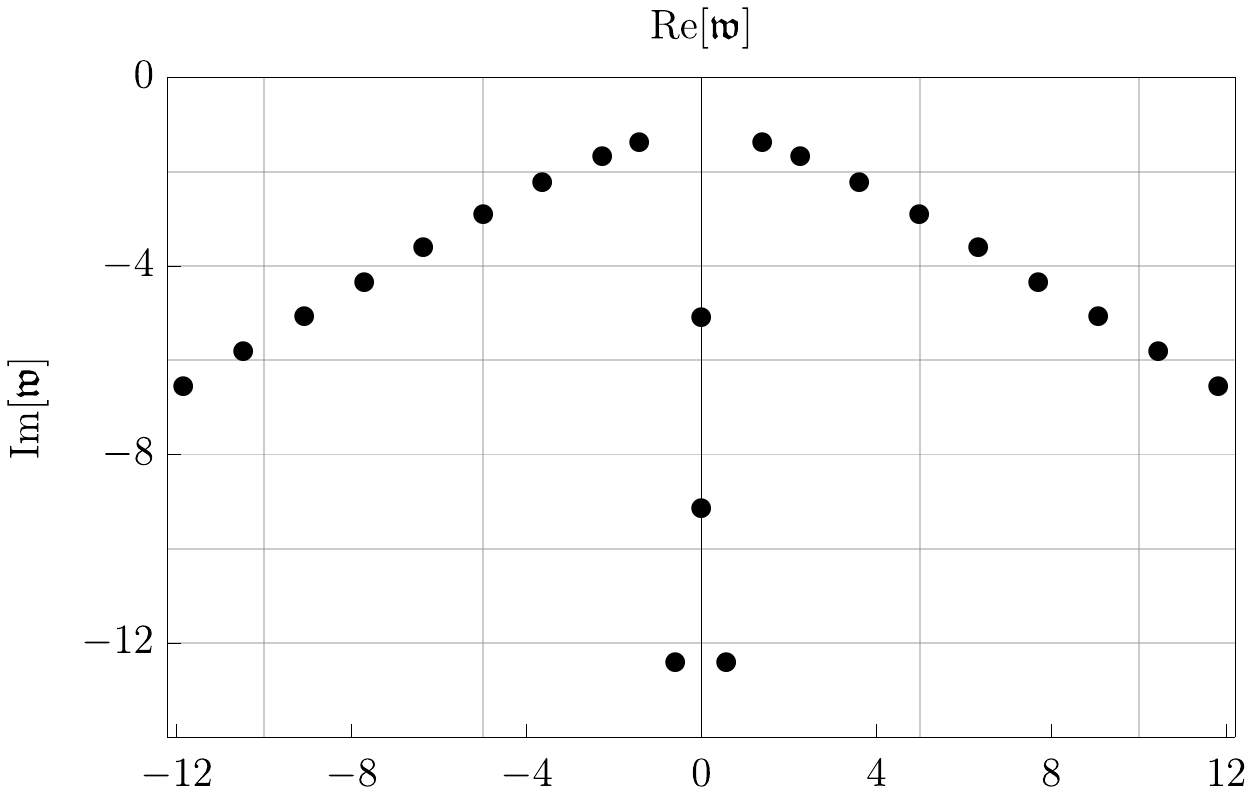}
\caption{The scalar channel spectrum at $\qfr = 0.5$, plotted for $\alpha = 0.02$ and $\gamma_2 = \{0.005, 0.0065, 0.0067, 0.007 \}$,  in progression from top-left, top-right to bottom-left and bottom-right.}
\label{fig:Ord2-Coll}
\end{figure*}

\begin{figure*}[h]
\centering
\includegraphics[width=0.45\textwidth]{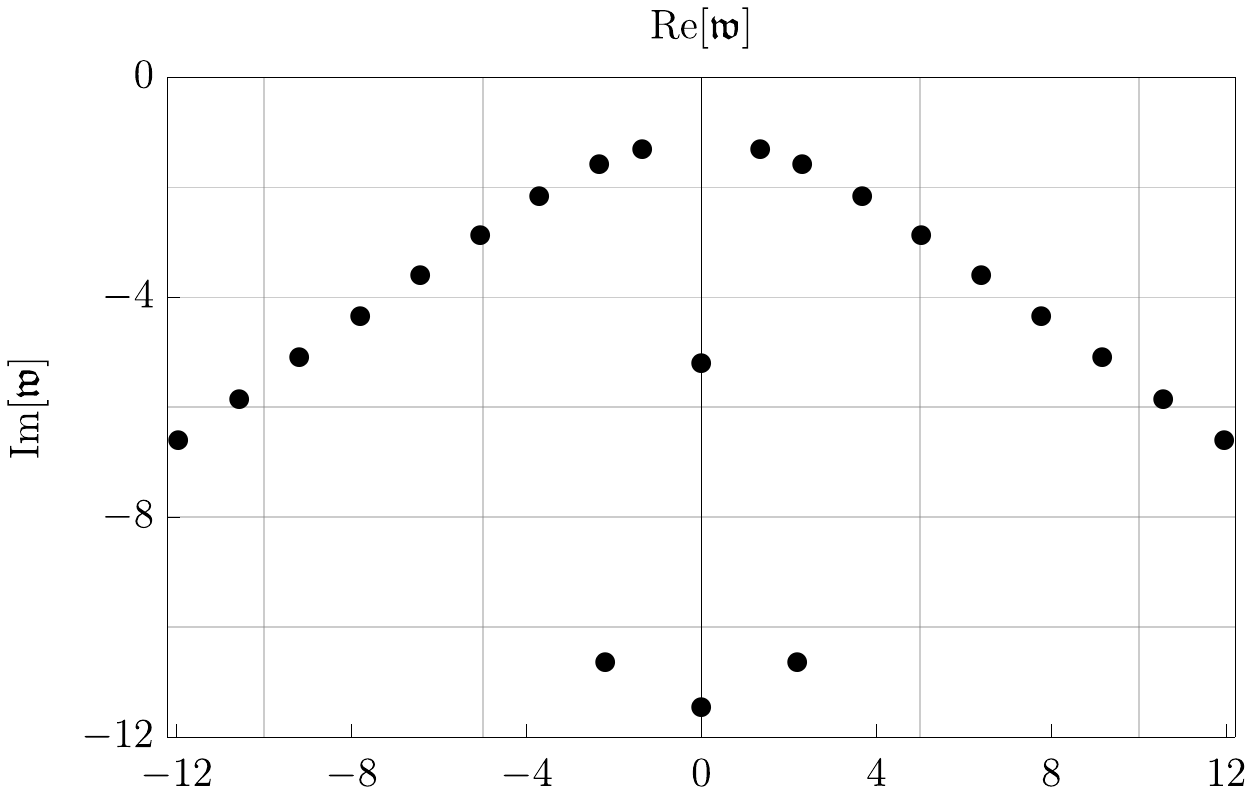}
\hspace{0.05\textwidth}
\includegraphics[width=0.45\textwidth]{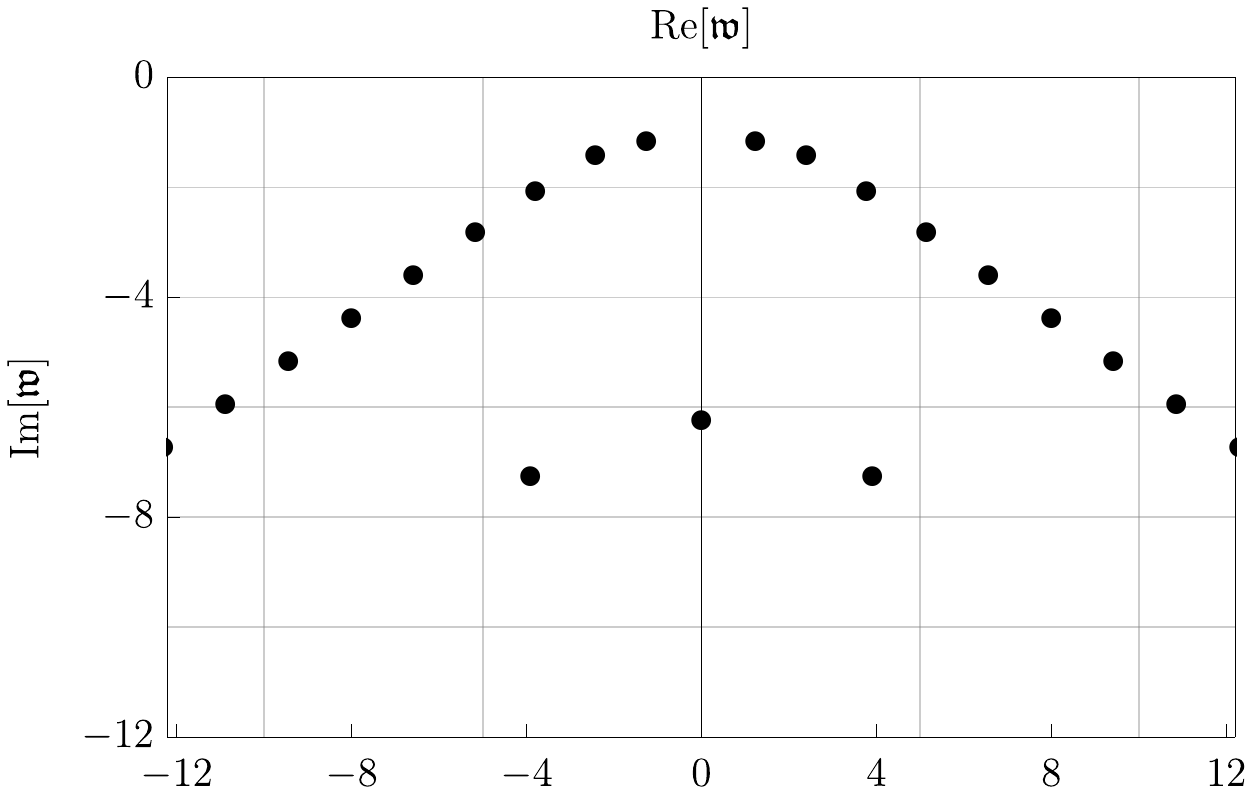}
\\
\includegraphics[width=0.45\textwidth]{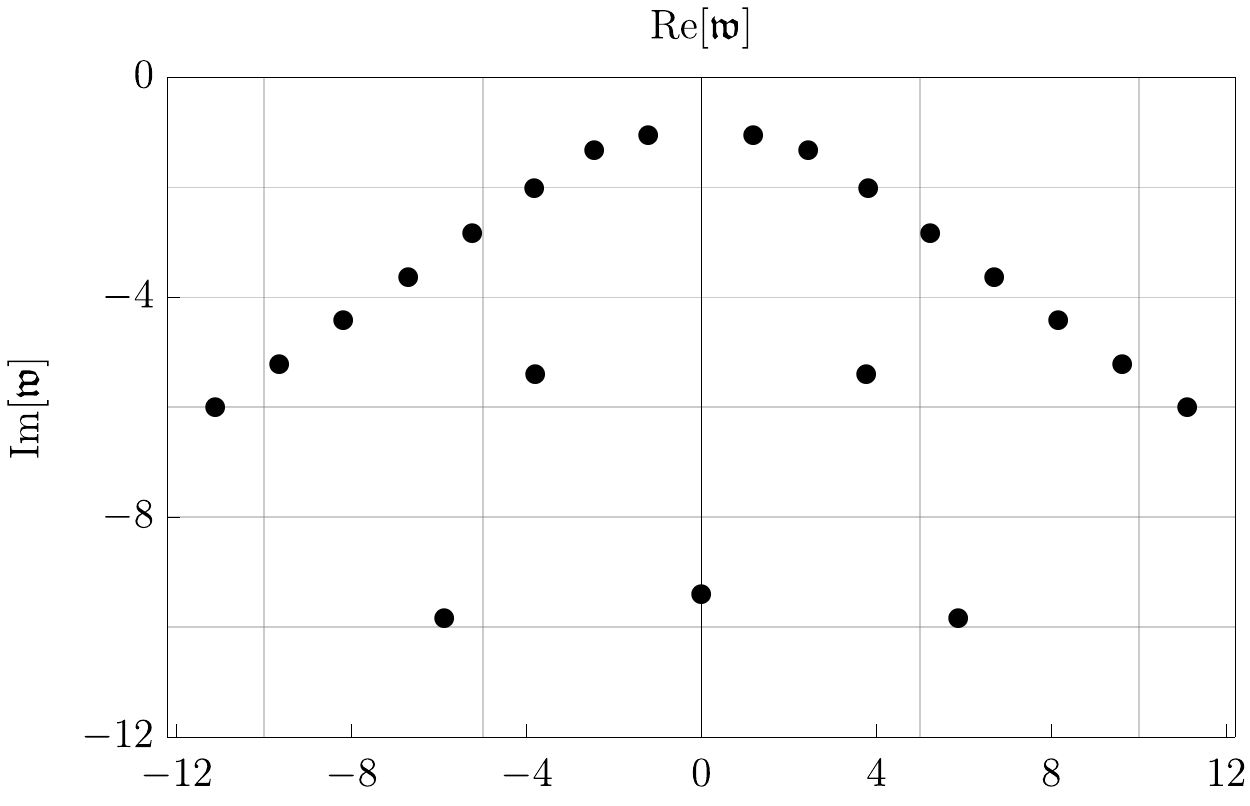}
\hspace{0.05\textwidth}
\includegraphics[width=0.45\textwidth]{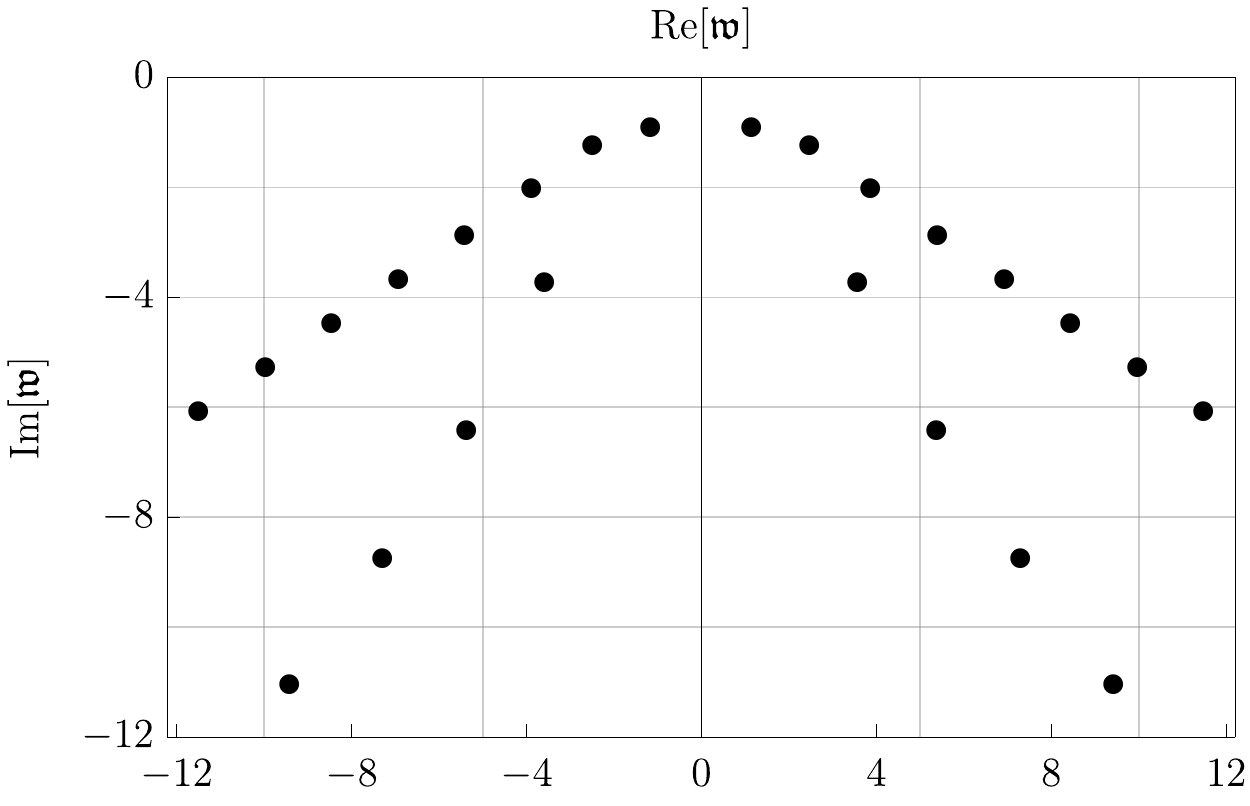}
\caption{The scalar channel spectrum at $\qfr = 0.5$, plotted for $\alpha = 0.02$ and 
$\gamma_2 = \{0.01, 0.02, 0.03, 0.05 \}$,  from top-left, top-right to bottom-left and bottom-right.}
\label{fig:Ord2-TwoBranches1}
\end{figure*}

A new phenomenon arises at order $O(\alpha^2)$, i.e. when $\gamma_1 \neq 0$ and $\gamma_2\neq 0$ in the action  \eqref{ActR2R4}. Here, in addition to the standard ``Christmas tree'' branches, the spectrum again exhibits multiple 
new poles on the imaginary axis  (Fig.~\ref{fig:Ord2-Coll}, top-left panel). The distance between the new poles depends on the 
ratio $\gamma_2/\gamma_1$. Changing this ratio, one observes pair-wise collisions of the poles (Fig.~\ref{fig:Ord2-Coll}, top-right and bottom-left panels), leading to them ``splitting off'' the imaginary axis (Fig.~\ref{fig:Ord2-Coll}, bottom-right panel). These multiple collisions  eventually result in forming a new pair of symmetric branches in the ``Christmas tree'' spectrum, as shown in 
  Fig.~\ref{fig:Ord2-TwoBranches1}. This behaviour appears to be robust with respect to changing the parameters of the action \eqref{ActR2R4}, as well as changing the value of the spatial momentum  $\qfr$.

The bottom-right panel of Fig.~\ref{fig:Ord2-TwoBranches1} is, in a sense, the main result of this paper: it demonstrates how new branches of poles in the quasinormal spectrum (expected in the dual QFT at finite coupling) can emerge from the higher-derivative terms in the gravitational action. We note that tuning the ratio $\gamma_2/\gamma_1$ should not be thought of as changing the coupling constant of the dual theory, but rather as changing the strength of subleading coupling-dependent corrections. From the point of view of supergravity, parameters such as $\gamma_1$ and $\gamma_2$ are determined by the  graviton scattering amplitudes, while we think of $\alpha$ as being the analog of $\alpha'$ in string theory.

\begin{figure*}[h]
\centering
\includegraphics[width=0.45\textwidth]{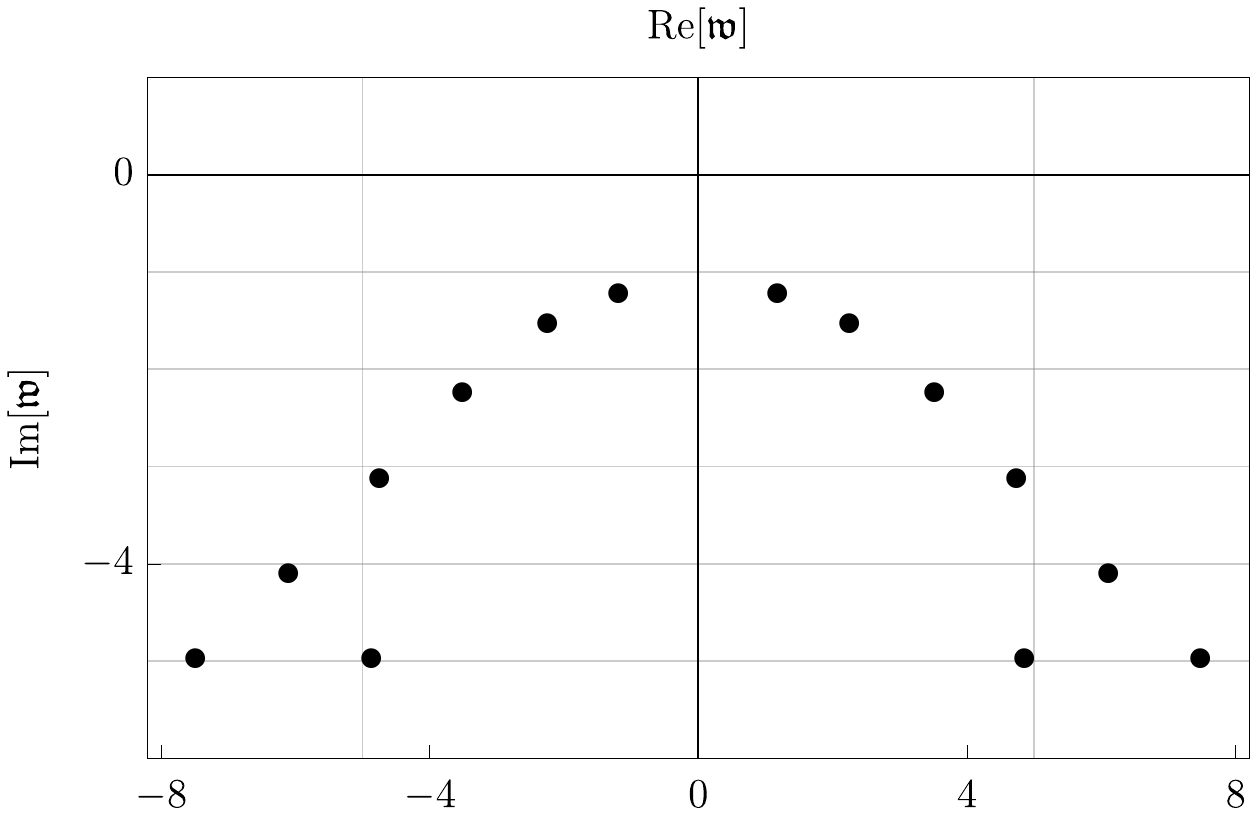}
\hspace{0.05\textwidth}
\includegraphics[width=0.45\textwidth]{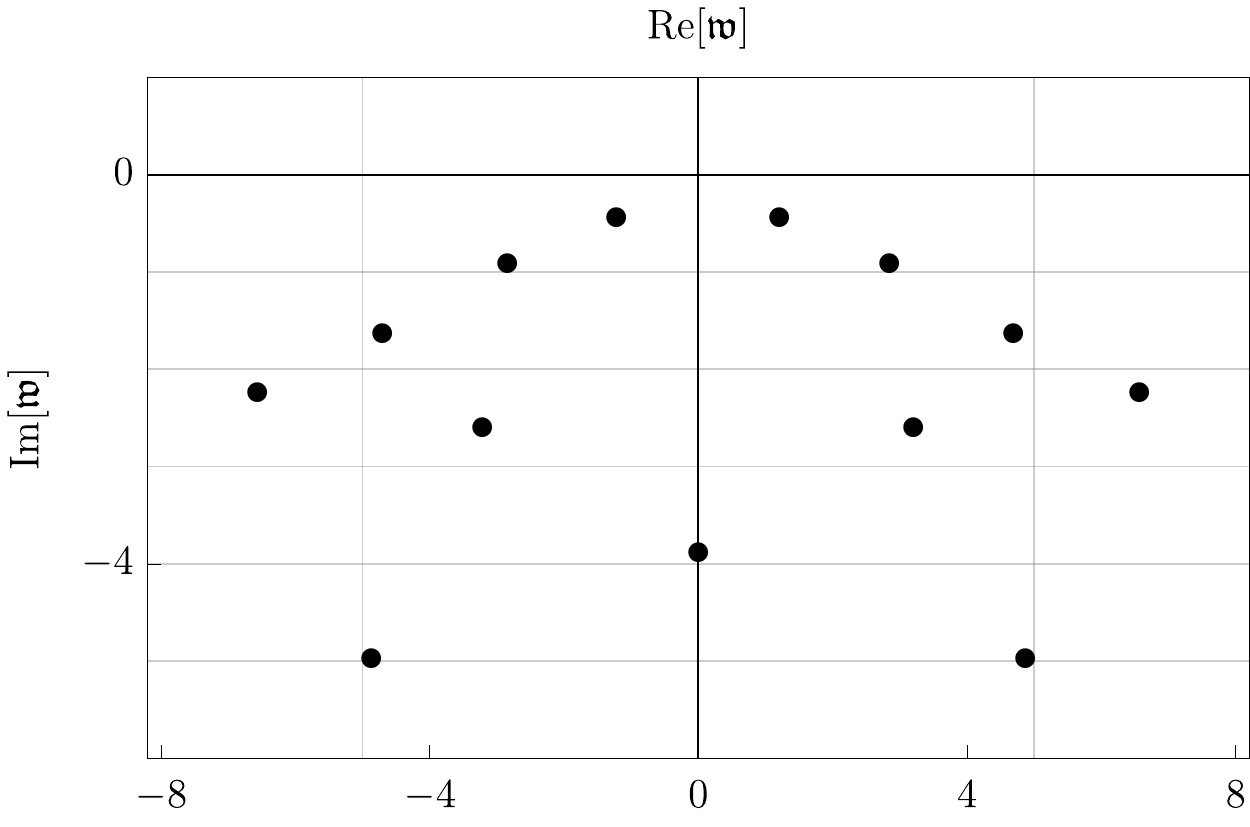}
\\
\includegraphics[width=0.45\textwidth]{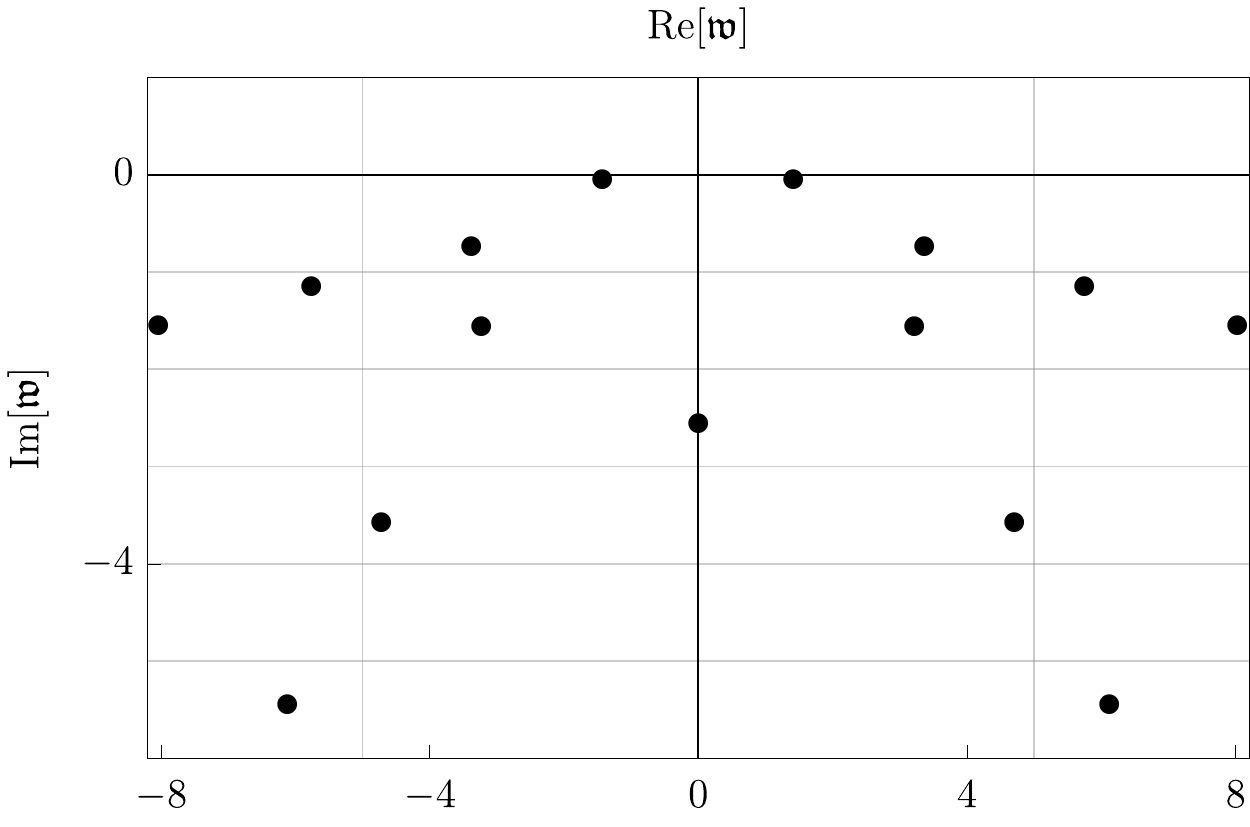}
\hspace{0.05\textwidth}
\includegraphics[width=0.45\textwidth]{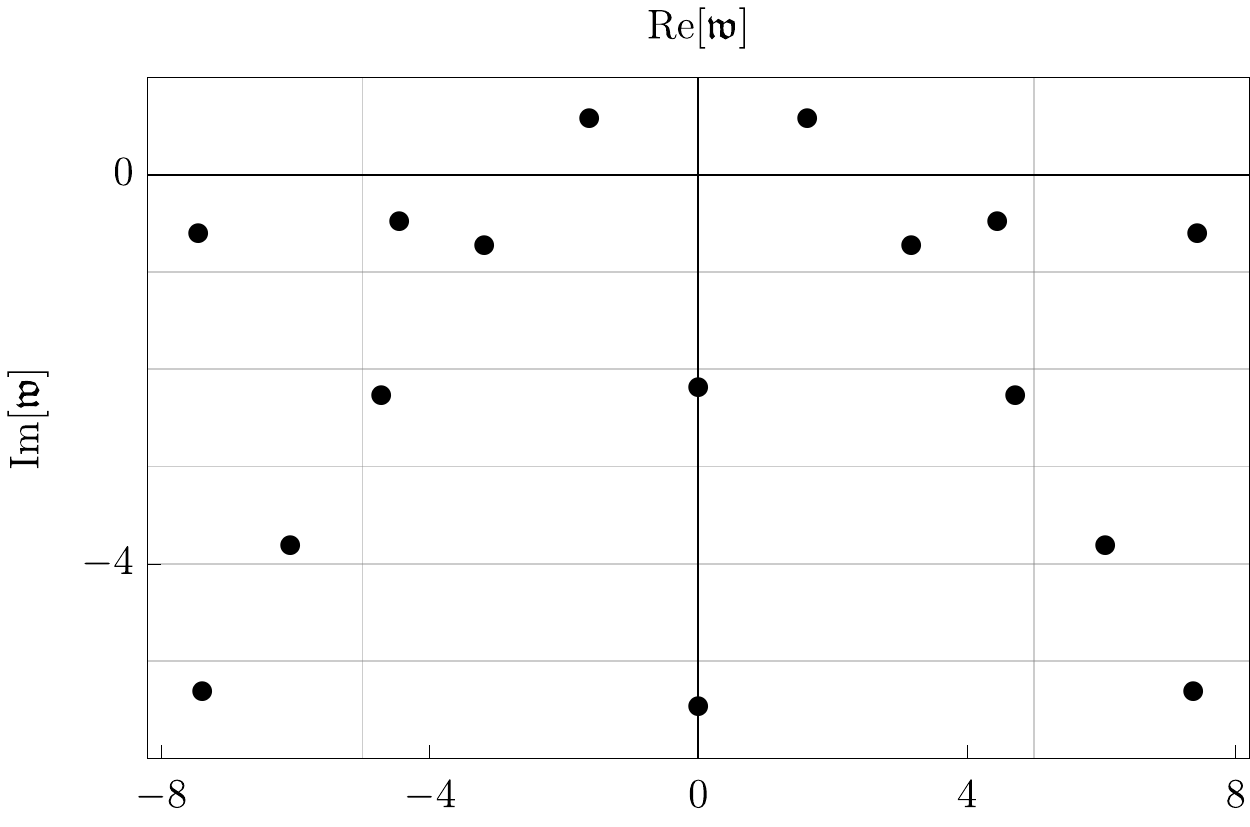}
\caption{The scalar channel  spectrum  at  $\qfr = 0$, plotted for $\alpha = \{0.01,0.04,0.06,0.08\}$ and $\gamma_2 = 0.05 $, from top-left, top-right to bottom-left and bottom-right.}
\label{fig:Ord2-TwoBranchesAlphaDepInst1}
\end{figure*}

\begin{figure*}[h]
\centering
\includegraphics[width=0.5\textwidth]{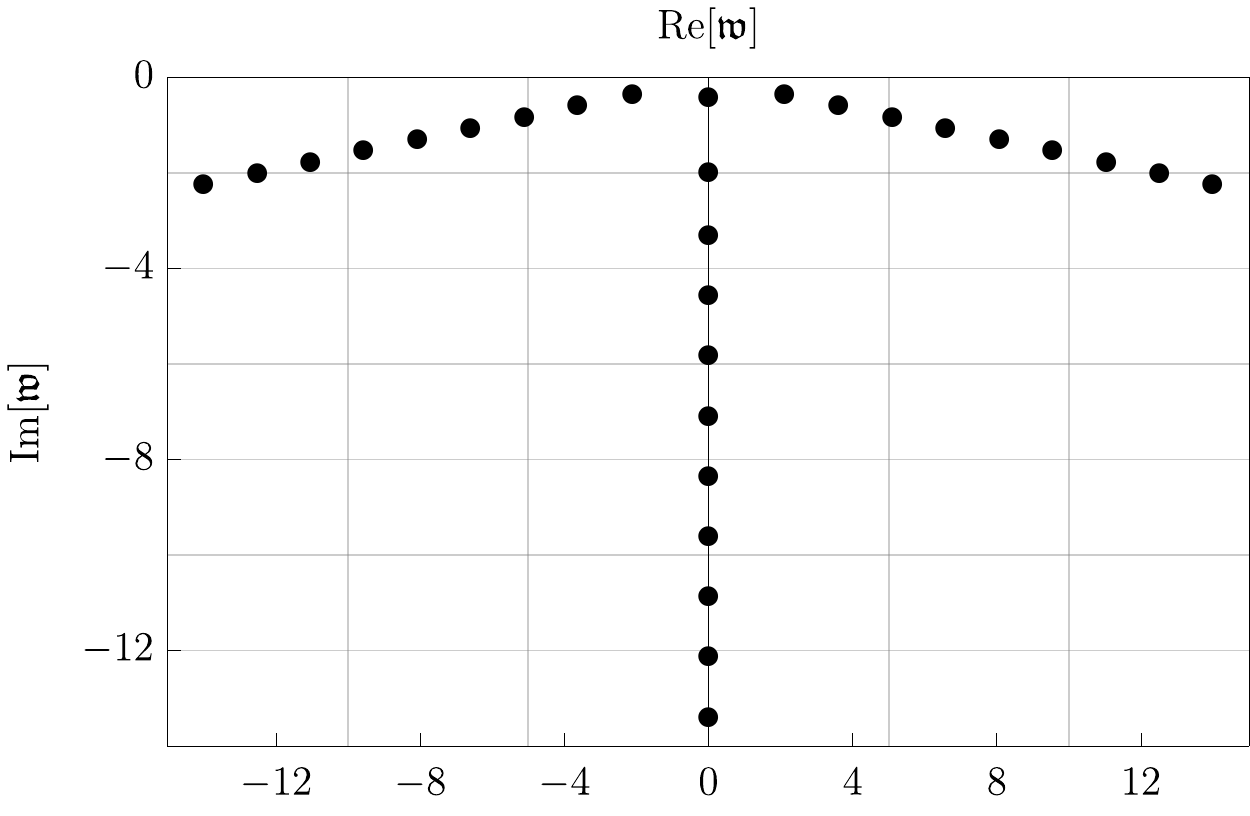}
\caption{The scalar channel quasinormal spectrum of the $O(\alpha^2)$ theory with $\qfr = 0$, plotted for $\alpha = 0.02$ and $\gamma_2 = -0.3$.}
\label{fig:Ord2-NegativeGamma}
\end{figure*}

We have also observed two other, qualitatively different, regimes that the spectrum can exhibit. One is shown in Fig.~\ref{fig:Ord2-TwoBranchesAlphaDepInst1}, where we fix the parameter $\gamma_2$ at the same value as in the bottom-right panel of Fig.~\ref{fig:Ord2-TwoBranches1}, $\gamma_2 = 0.05$,  and change $\alpha$ at $\qfr=0$.  Here, for sufficiently large $\alpha$, the second pair of branches formed by the same mechanism of the imaginary axis pole collisions, ``goes up and through''  the first pair of branches, eventually crossing into the upper half-plane of complex frequency thus developing a fatal instability. This simply means that not all ratios of the parameters in the action \eqref{ActR2R4} are physically allowed, assuming the action serves as a dual gravity description of a stable thermal QFT. Yet another pattern of the quasinormal spectrum following from the action \eqref{ActR2R4} is shown in Fig.~\ref{fig:Ord2-NegativeGamma}. For negative $\gamma_2$ and $\qfr=0$, we observe the emergence of a large number of poles on the imaginary axis without the formation of the two extra branches of the ``Christmas tree''. 
 %
%


\section{Thermal retarded correlators at $\qfr =0$}
\label{sec:zero-q}
At zero coupling, finite-temperature retarded two-point functions $G^R(\wfr,\qfr=0)$ in $4d$ have poles rather than branch cuts in the complex frequency plane. For example, for bosonic degrees of freedom, we have  \cite{Hartnoll:2005ju} 
\begin{align}\label{zeroq}
 G^R (\wfr) \sim \wfr^4 \psi \left( -i \wfr/2 \right) + P(\wfr),
\end{align}
where $\psi(z)$ is again the logarithmic derivative of the Gamma function and $P(z)$ is a polynomial 
of degree less than four. 
\begin{figure}[ht]
\centering
\includegraphics[width=0.45\textwidth]{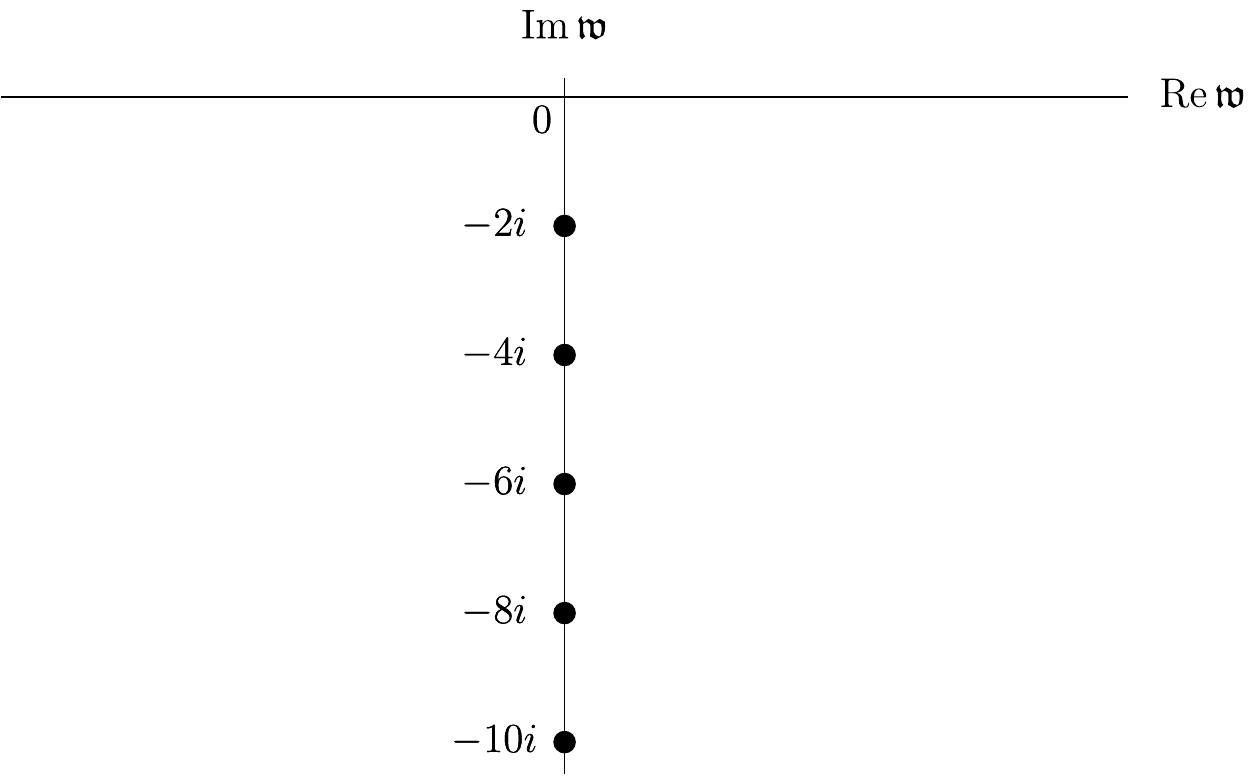}
\includegraphics[width=0.45\textwidth]{qnm_diagram_poles.pdf}
\caption{Singularities of a thermal retarded two-point  function $G^R(\wfr,\qfr)$ at $\qfr = 0$ in the complex frequency plane at zero coupling \cite{Hartnoll:2005ju} (left panel) and infinitely large coupling \cite{Starinets:2002br} (right panel).}
\label{fig:cuts_poles_q=0}
\end{figure}
Singularities of the correlator \eqref{zeroq} are simple poles located on the imaginary axis with periodicity $|\wfr|=2$, as shown in the left panel of Fig.~\ref{fig:cuts_poles_q=0}.  At infinite coupling, the 
same correlator has poles that are still arranged in the ``Christmas tree'' pattern, similar, although not identical to the $\qfr\neq 0$ spectrum (Fig.~\ref{fig:cuts_poles_q=0}, right panel).

When a higher-derivative term is added to the dual gravity action (one may consider e.g. the action   \eqref{ActR2R4} with $\gamma_1\neq 0$ and $\gamma_2=0$), the 
spectrum changes similarly to what is shown in  the right panel of  Fig.~\ref{fig:cuts_poles_finite}: the two branches of poles move up and become more dense (and interlaced with zeros), and new poles appear on the imaginary axis. Adding another higher-derivative term to the action with the right coefficients can result in the spectrum shown in Fig.~\ref{fig:Ord2-NegativeGamma}. We may speculate that in the limit of vanishing coupling, the two branches of the original ``Christmas tree'' form a branch cut along the entire real $\wfr$-axis (rather than the two cuts at $\qfr\neq 0$ shown in the left panel of Fig.~\ref{fig:cuts_poles}), with vanishing discontinuity and the only singularity remaining at infinity. Such a branch cut is a contractible cycle on the Riemann sphere, which can be ignored. Furthermore, the poles on the imaginary axis are re-arranged to form the pattern shown in the left panel of Fig.~\ref{fig:cuts_poles_q=0}. 

\section{Discussion}
We have seen that adding generic, modulo field redefinitions, higher-derivative terms to the Einstein-Hilbert gravity 
action can result in the emergence of new branches of quasinormal modes in the spectrum of black branes. This is qualitatively similar to the expected behaviour of dual thermal QFTs at finite coupling. Although our discussion has been necessarily qualitative and descriptive throughout, we believe it is important to point out that the potential mechanism of reconciling the two panels of  Fig.~\ref{fig:cuts_poles} exists in holography. 

The parameter space of the higher-derivative coefficients we are considering in this note is two-dimensional. Quantitatively, even in this simple case, we seem to have a continuum of possibilities for the resulting spectrum, yet, qualitatively, we observe only three standard patterns according to which the poles arrange themselves.  If the emerging qualitative picture is indeed correct, and assuming monotonicity of the coupling dependence, this implies that the coefficients of the higher derivative terms in the dual  action must lie within certain highly constrained intervals in order for the finite-temperature theory to have an expected behaviour at weak 
 coupling. This is similar to the constraint on the ratio of shear viscosity to entropy density emerging from the expected behaviour in kinetic theory \cite{Grozdanov:2016vgg}. 

It would be highly desirable to understand the analytic structure of  real-time finite-temperature correlators of composite operators at weak but non-vanishing coupling beyond the results obtained in refs.~\cite{Moore:2018mma,Romatschke:2015gic, Kurkela:2017xis,Grozdanov:2018atb}.

\acknowledgments{S. G. was supported by the U. S. Department of Energy under grant Contract Number DE-SC0011090. S. G. would also like to thank the Centre for Theoretical Physics (CPhT) at \'{E}cole 
Polytechnique and particularly Blaise Gout\'{e}raux for hospitality, and CNRS for financial support during this visit.}

\bibliographystyle{JHEP}
\bibliography{Genbib}{}
\end{document}